\documentclass[a4paper,10pt]{article}

\usepackage{cite}
\usepackage{color}
\usepackage[usenames,dvipsnames]{xcolor} 
\definecolor{DarkGreen}{RGB}{0,64,0}

\usepackage[]{array}
\usepackage[]{amsmath}
\usepackage[]{amssymb}
\usepackage[]{mathrsfs}
\usepackage[]{tensor}
\usepackage{bm}

\begin{document}
\vskip 2cm

\begin{center}

{\LARGE{\bf Small black holes in the large $\boldsymbol{D}$ limit}}
\vskip 1cm

{\Large Predrag Dominis Prester}\\
{}~\\
\quad \\
{\em Department of Physics, University of Rijeka,}\\[1mm]
{\em  Radmile Matej\v{c}i\'{c} 2, HR-51000 Rijeka, Croatia}\\
\vskip 1cm
Email: pprester@phy.uniri.hr

\end{center}

\vskip 2cm 

\noindent
{\bf Abstract.}  

\bigskip

The large $D$ limit of AdS$_2 \times S^{D-2}$ solutions in the particular higher-derivative Lovelock-type theory is analyzed. The theory and the solutions were originally considered in an attempt to effectively describe near-horizon behavior of $D$-dimensional spherically symmetric 2-charge small extremal black holes which in superstring theory context are assumed to correspond to configurations in $S^1 \times T^{9-D}$ compactification schemes in which fundamental string is wound around circle $S^1$. Though in $D\to\infty$ limit the action contains infinite number of higher-derivative terms, their contributions to equations of motion sum into simple exponential form which allows us to find explicit solutions. A simplicity of this example gives us the opportunity to study some connections between $\alpha'$ and $1/D$ expansions. In the leading order in $1/D$ the relation between the string parameter $\alpha'$ and the radius of the horizon 
$r_{\mathrm{h}}$ (in the string frame) satisfies $r_{\mathrm{h}} \propto D \sqrt{\alpha'}$, i.e., we obtain an explicit realization of the relation inferred by Emparan \emph{et al.} in the different context of large black holes in the ordinary Einstein gravity where $\alpha'$ is not manifestly present.

\vskip 1cm 

 
\vfill\eject

\section{Introduction and summary of the results}
\label{sec:intro}

\bigskip

The idea of studying gravity in the limit of large number of spacetime dimensions $D$ dates back at least to \cite{Strominger:1981jg}. Original attempts were focused to quantization of gravity, motivated by the possible analogy with large $N$ limit of $SU(N)$ gauge theories 
\cite{Strominger:1981jg,BjerrumBohr:2003zd,Hamber:2005vc}. Though this original application so far did not meet the expectations, it was shown that the idea may be useful in other contexts 
\cite{Soda:1993xc,Yoshino:2002br,Kol:2004pn,Asnin:2007rw,Caldarelli:2008mv,Canfora:2009dx,%
Hod:2010zza,Hod:2011zza,Hod:2011zzb,Hod:2011zzc,Coelho:2012sya,Caldarelli:2012hy}. 

Recently a new fresh blood was injected to the idea. A detailed study of black hole solutions in 
classical General Relativity in the large $D$ limit was initiated in \cite{Emparan:2013moa}. It was shown that General Relativity in $D \to \infty$ limit drastically simplifies, which makes a good starting point for a perturbative expansion in $1/D$. The hope is that this expansion may be useful for answering some realistic questions in $D=4$. In the follow-up paper \cite{Emparan:2013xia} it was shown that in $D \to \infty$ limit of GR a broad class of (large) neutral black holes has a universal near-horizon limit, with the limiting geometry being the two dimensional black hole of string theory with a two dimensional target space. To achieve a connection with string geometry, one needs to make identification
\begin{equation} \label{Daprime}
\sqrt{\alpha'} \sim \frac{r_0}{D}
\end{equation}
where $r_0$ is the radius of the horizon and $\alpha'$ is the string parameter. In \cite{Giribet:2013wia} similar large $D$ limit was applied to the special class of higher-derivative Lovelock gravity theories.

In this letter we want to study large $D$ limit of AdS$_2 \times S^{D-2}$ backgrounds which describe near-horizon region of spherically symmetric ``small'' extremal black holes, i.e., zero-temperature black holes whose horizons are resolved by higher-curvature terms present in the  action. In the string theory context, small black holes are assumed to describe ``stringy'' objects with the size of the order of the string length $l_s = \sqrt{\alpha'}$. As in some regimes effective string coupling constant is small, such objects have potential to offer us a window into small distance behavior of gravity without worries about effects of quantum corrections on the tree-level result. However, there is a problem with this program - because spacetime curvature in the near-horizon region is of the order of 
$1/\alpha'$ low-energy expansion is expected to break down and one cannot trust low-energy string effective actions for studying such backgrounds. Indeed, except for some 4-dimensional cases, low-energy effective actions are producing wrong results for the entropy of such small extremal black holes.  For the development and status of the current understanding of small extremal black holes in the heterotic string theory see, e.g., 
\cite{Dabholkar:2004yr,Dabholkar:2004dq,Sen:2004dp,Hubeny:2004ji,Sen:2005kj,Prester:2005qs,%
LopesCardoso:2006bg,Cornalba:2006hc,Giveon:2006pr,Dabholkar:2006tb,Castro:2007hc,Cvitan:2007en,%
Lapan:2007jx,Alishahiha:2007nn,Cai:2007cz,Sen:2009bm}  (for a more complete set of references one can consult reviews \cite{Sen:2007qy,Larsen:2008qn,Prester:2010cw}). In the following we shall take standard assumption that effective space-time near-horizon description for heterotic string configurations of our interest is possible, and in particular that near-horizon geometry is AdS$_2 \times S^{D-2}$.

To study large $D$ limit of small black holes in the classical regime of gravity theory, while also keeping the contact with the string theory, we propose to start with a simple and accessible example. Such example is provided by small spherically symmetric extremal 2-charge black holes in the particular ``Lovelock-type'' higher-curvature action, whose near-horizon AdS$_2 \times S^{D-2}$ solution was constructed in \cite{Prester:2005qs}.\footnote{To avoid possible confusion, let us emphasize that Lovelock theories in \cite{Giribet:2013wia} are different from those we use in this paper.} This action was invented in an attempt to provide effective near-horizon description of the supersymmetric black hole configurations in the heterotic string theory compactified on $S^1 \times T^{9-D}$, where $4 \le D \le 9$, which correspond to fundamental string wound $w$ times around $S^1$ which has momentum in the direction of $S^1$ characterized by the natural number $n$. Indeed, it was shown in \cite{Prester:2005qs} that this rather simple effective action provides near-horizon extremal black hole solutions with the Wald entropy matching statistical entropy (obtained by microstate counting inside the string theories) at the lowest order in the string coupling $g_s$ (but exactly in $\alpha'$). This Lovelock-type action is not equal to the tree-level effective low-energy action of the heterotic string theory in any $D$, so its meaning in the string theory context is not clear. In any case, it is at least an interesting toy model to study large $D$ limit. Though this particular stringy interpretation is meaningful only for $D \le 9$, the theory and solutions can be extended to all $D \ge 4$, which allows us to study formally the $D \to \infty$ limit. Also, in $D\to\infty$ limit this theory has infinite number of higher-derivative terms with unlimited order, which is a property of string low-energy effective actions (but already at finite $D$).   

Though in the course of the paper we frequently refer to string theory connection (and even borrow a notation), we would like to emphasize that our analysis is not restricted to this context and does not rest on the validity of this connection. Lovelock-type actions are interesting by themselves as alternative theories of gravity - they are higher-derivative gravity theories which lead to normal second-order Euler-Lagrange equations and which do not induce new degrees of freedom (ghosts) when perturbed around large classes of backgrounds. Our AdS$_2 \times S^{D-2}$ configurations are exact classical solutions in one such theory, which makes them valuable topic of interest, even aside mentioned applications in string theory or in descriptions of small black holes. 

Our study has an important difference from \cite{Emparan:2013xia}. The black holes we study are small, with the horizon of the size of the order of $\sqrt{\alpha'}$ (singular in the lowest, i.e., two-derivative, order), in which contributions to near-horizon properties of all higher-derivative terms are of the same order in $\alpha'$.\footnote{Formally, $\alpha'$ is the parameter setting the scale for the higher-curvature terms in the action, which in string theory interpretation is equal to the string parameter $\alpha'$ and so we chose to use the same symbol.} So, in our case $\alpha'$ is already present from the start and we are interested in the ``interplay'' between $D$ and $\alpha'$ (and possibly other parameters).

Let us summarize our main results here: (i) We find that AdS$_2 \times S^{D-2}$ solutions of our Lovelock-type theory take remarkably simple form in $D \to \infty$ limit. Though higher-derivative terms of all orders contribute to equations, this infinite sum takes a simple form of the exponential function. This allowed us to find simple analytic solutions, which was not the case for finite $D>7$. Our results support the expectations that large $D$ limit may be a useful tool, at least for classical analyzes. (ii) We show that the relation between $\alpha'$ and $D$ in $D \to \infty$ limit is consistent with the relation (\ref{Daprime}) from \cite{Emparan:2013xia}. More precisely, it takes the form
\begin{equation} \label{Dapexact}
\sqrt{\alpha'} = c\, \frac{r_{\mathrm{h}}}{D} \;, \qquad c = 2 \sqrt{\ln{2}}
\end{equation}
where $r_{\mathrm{h}}$ is the proper radius of the black hole horizon in the string frame metric (all our statements here refer to the string frame). (iii) All Euler densities $E_k$ (where $k$-order density has 
$2k$-derivatives) have the same leading order at large $D$, which is $O(1)$ (they are finite in $D\to\infty$ limit). A similar statement holds for $\alpha'$-expansion (expansion in derivatives) where all higher curvature terms multiplied by the corresponding power of $\alpha'$ (i.e., $\alpha'^k R^k$, which is what one has in the effective action) evaluated on near-horizon extremal small black hole solution are $O(1)$ in $\alpha'$. We see here explicit realization of the connection between $\alpha'$ and $1/D$ expansions hinted in \cite{Emparan:2013xia}. However, in our example we can spot a notable difference - in $S^{D-2}$ spacetime block only Ricci scalar factors contribute in the leading order (contractions containing Ricci tensor and/or Riemann tensor are subleading in $1/D$, though they are of the same order in $\alpha'$). (iv) Conclusions (i)-(iii) are not generic, e.g., they are not satisfied if we truncate the action to fixed finite order in higher-derivative expansion while taking $D \to \infty$. We demonstrate this on an example of 4-derivative theory in which higher-derivative part of the Lagrangian contains only second Euler density (so called Gauss-Bonnet term). In this theory instead of (\ref{Daprime}) we have $\alpha' \sim 1/D$. The fact that theory contains infinite number of higher-derivative terms of unlimited order when $D\to\infty$ appears to crucially affect its large $D$ properties.

The outline of the paper is as follows. In Section \ref{sec:lovelock} basic properties of Lovelock-type theory and its two charge AdS$_2 \times S^{D-2}$ solution (for finite $D$) are reviewed, together with its connection with small extremal black holes and heterotic string theory. In Section \ref{sec:Dtoinfty} we perform large $D$ limit on this solution and analyze its properties. In Section \ref{sec:finite} we perform the same calculations inside the theory obtained by truncating the higher-derivative sector to 4-derivative part and show that the solution in large $D$ limit has markedly different properties from the corresponding one from Section \ref{sec:Dtoinfty}.

\vspace{20pt}

\section{Lovelock-type action and small black holes}
\label{sec:lovelock}

\bigskip

In \cite{Prester:2005qs} we introduced the following higher-derivative action
\begin{equation} \label{treeea}
\mathcal{A} = \mathcal{A}_0 + 
 \frac{1}{16\pi G_D} \sum_{k=2}^{\infty} \lambda_k \int d^Dx \sqrt{-g} \,e^{-\Phi} \, E_k \,,
\end{equation}
where $E_k$ are Euler densities (or extended Gauss-Bonnet densities)
\begin{equation}\label{lgbm}
E_k = \frac{1}{2^k}
\, \delta_{\mu_1\nu_1\ldots\mu_k\nu_k}^{\rho_1\sigma_1\ldots
\rho_k\sigma_k} \, {R^{\mu_1\nu_1}}_{\rho_1\sigma_1}\cdots
{R^{\mu_k\nu_k}}_{\rho_k\sigma_k}\;,
\end{equation}
and $\mathcal{A}_0$ denotes lowest-order (2-derivative) part given by
\begin{equation} \label{eaDs}
\mathcal{A}_0 = \frac{1}{16\pi G_D} \int d^D x \, \sqrt{-G} \, e^{-\Phi} \, 
\left[ R + (\partial_\mu \Phi)^2  -  T^{-2} \, (\partial_\mu T)^2
 - T^2 \, (F^{(1)}_{\mu\nu})^2 - T^{-2} \,  (F^{(2)}_{\mu\nu})^2 \right] \, ,
\end{equation}
Here $\Phi$ and $T$ are scalar fields, $F^{(1)}$ and $F^{(2)}$ two $U(1)$ gauge fields, and $G_D$ is
the Newton constant. The coefficients $\lambda_k$ in (\ref{treeea}) are given by
\begin{equation}\label{coup}
\lambda_k = \frac{4\, \alpha'^{k-1}}{4^k k!} \,, 
\end{equation}
where $\alpha'$ is a constant which in stringy interpretation becomes (squared) string length parameter. The theory defined by (\ref{treeea})-(\ref{coup}) can be defined in any number of spacetime dimensions 
$D$. Though we left the upper bound in the sum in (\ref{treeea}) unlimited, with intent to show that the action has the same form in all $D$, it is well-known that for $k > [D/2]$ Euler densities vanish. So the upper limit of the sum can be put to $[D/2]$ and we see that there are finite number of terms in the action (having up to $2[D/2]$ derivatives). 

Let us first neglect higher-curvature part of the action. Then we are left with the action $\mathcal{A}_0$ which is known to be the bosonic part of a particular $N=2$ supergravity. This theory has supersymmetric (BPS) solutions with the geometries of asymptotically flat spherically symmetric extremal black holes \cite{Peet:1995pe}. However, these black holes have singular horizons which have vanishing area. When higher-curvature terms are switched on (by taking $\alpha' \ne 0$ in (\ref{treeea})) one expects that black holes get ``regularized'' - the horizon becoming regular, with nonvanishing area and the radius of the order $\sqrt{\alpha'}$. Such black holes are referred to as small. Assuming that these black holes are still extremal ($T_{bh}=0$) one expects that the near-horizon geometries are $AdS_2 \times S^{D-2}$.

For $D$ satisfying $4 \le D \le 9$ the action $\mathcal{A}_0$ (\ref{eaDs}) is equal to the lowest-order (2-derivative) bosonic part of the effective action of the heterotic string theory compactified on $S^1 \times T^{9-D}$, consistently truncated to a sector in which the only nonvanishing fields beside metric are Kaluza-Klein fields coming from $S^1$ compactification: two gauge fields  and one scalar modulus field $T$. In 2-charge black hole solutions of $\mathcal{A}_0$ mentioned above, one charge is proportional to the winding number $w$ of the fundamental string on the compactification circle $S^1$, while the other charge is proportional to the momentum number $n$ along the same $S^1$. BPS condition imposes $nw > 1$. One can calculate the statistical entropy of such BPS states, which are characterized by fixed $n$ and $w$ with $nw > 0$, by direct counting of microstates, and the result in the leading order in $nw$ is
\begin{equation} \label{Smicro}
S_{\mathrm{micro}} = 4\pi \sqrt{nw} \;, \qquad nw \gg 1
\end{equation}

When one tries to use full heterotic low-energy effective action to find regulated small black hole solutions, one immediately hits the following problem - due to the fact that curvature around the horizon goes as $1/\alpha'$, all higher-derivative terms contribute in same order in $\alpha'$ so the low-energy effective expansion breaks down. Even the techniques of summing up all $\alpha'$-corrections, which are working for large extremal black holes in $D=4$ and $D=5$, are not producing correct results for black hole entropy (\ref{Smicro}) in the small black hole limit in $D=5$ \cite{Castro:2007hc,Cvitan:2007en,Prester:2008iu,Prester:2010cw}. A proper construction of the near-horizon description for the 2 charge extremal small black holes in heterotic string theory is an interesting issue which is still not completely settled.

It was our goal in \cite{Prester:2005qs} to try to use the relatively simple Lovelock-type action (\ref{treeea}), as a sort of a toy theory,\footnote{Now, one can still be puzzled how a simple action like (\ref{treeea}) can ``substitute'' full heterotic low-energy effective action (HLEEA) which is very different and much more complicated (see., e.g., \cite{Gross:1986mw,Liu:2013dna}). In particular: (A) HLEEA contains infinite number of higher-derivative terms, organized as expansion in string parameter $\alpha'$, some of which contain covariant derivatives and gauge fields, while higher-derivative terms in (\ref{treeea}) are finite in number and contain just the metric and no covariant derivatives; (B) HLEEA does not contain $\alpha'^2$ (6-derivative) terms which are field-redefinition invariant, while (\ref{treeea}) contain such terms inside Euler density term $E_3$; (C) HLEEA contain gravitational Chern-Simons terms, while in (\ref{treeea}) there are no such terms. Part of the explanation for (A) lies in $SO(1,1) \times SO(D-2))$ symmetry of the solutions we consider here, which makes all terms which include covariant derivatives irrelevant. Moreover, in case of large extremal geometries it was shown that in a particular scheme, due to the special property of solutions (``parallelizable torsion''), only finite number of terms in the action are relevant \cite{Prester:2008iu,Prester:2009mc,Prester:2010cw}. So, it is not unreasonable to expect that something similar happens for small geometries. To answer (B), we note that analysis from \cite{Prester:2005qs} implies that only relevant higher-derivative terms in (\ref{treeea}) are not invariant on field redefinitions. The objection (C) is the trickiest, especially because for large extremal geometries the relevant higher-derivative terms in HLEEA are exactly those originating from gravitational Chern-Simons term \cite{Prester:2008iu,Prester:2009mc,Prester:2010cw}. On this we just note that both gravitational Chern-Simons terms and Euler terms are closely connected with anomalies. It is important to remember that LEEA are by construction not appropriate for addressing small geometries, so use of (\ref{treeea}) is an attempt of a different type of an effective description. Let us also add here that there are also interesting similarities between HLEEA and (\ref{treeea}). For example, if we identify parameter $\alpha'$ in (\ref{treeea}) with the string tension, then the coefficient $\lambda_2 = \alpha'/8$ is the same in the two actions, and in general the coefficients $\lambda_k$ have dependence on $k$ which roughly corresponds to the behavior of coefficients multiplying Riemann$^k$-type terms in the heterotic actions.} to obtain reasonable near-horizon solutions with the correct expression for the black hole entropy in all $4\ge D\le 9$.\footnote{Note that we cannot claim that full (i.e., in the whole space) asymptotically flat extremal black hole solutions with $AdS_2 \times S^{D-2}$ near-horizon geometry indeed exist in the Lovelock-type theory. In \cite{Chen:2009rv} it was claimed, based on the numerical analysis, that in $D=4$ it indeed does not exist. However, this analysis is not conclusive because some 4-derivative Lagrangian terms were neglected in the calculation, so this result is not obtained for (\ref{treeea}). Beside, outside the near-horizon region there is no reason to believe that action (\ref{treeea}) can be used as effective description of heterotic string theory, so this question is meaningless in string theory context} 
Near-horizon geometry of extremal black holes in $D$ dimensions is expected to be isometric to 
$AdS_2 \times S^{D-2}$, which in the case of the theory (\ref{treeea})-(\ref{coup}) is
\begin{eqnarray} \label{nhsmall}
&& ds^2  = v_1 \left( -r^2 dt^2 + \frac{dr^2}{r^2} \right)  + v_2\, d\Omega_{D-2} \; ,
\nonumber \\ 
&& e^{-\Phi} = u_S \;, \quad T = u_T \;, \quad F^{(1)}_{rt} = e_1 \;, \quad F^{(2)}_{rt} = e_2  \;.    
\end{eqnarray}
where $d\Omega_k$ denotes standard metric on the unit $k$-dimensional sphere, and $v_{1,2}$,
$u_{S,T}$, and $e_{1,2}$ are constants to be determined from equations of motion. By using Sen's entropy function formalism \cite{Sen:2005wa,Sen:2007qy} we found \cite{Prester:2005qs}
\begin{eqnarray} \label{sollove}
&& v_1 = \frac{\alpha'}{2} \;, \qquad  v_2 = v_2(D)  \;, \qquad
 u_S = \frac{8\pi G_D}{\Omega_{D-2}} \frac{\sqrt{|nw|}}{v_2^{(D-2)/2}} \;,
\nonumber \\ 
&& u_T = \sqrt{\left| \frac{n}{w} \right|} \;, \qquad
 e_1 = \frac{\sqrt{\alpha' |nw|}}{2n} \;, \qquad
 e_2 = \frac{\sqrt{\alpha' |nw|}}{2w} \;.
\end{eqnarray}
Normalized electric charges $n$ and $w$, which in the heterotic string theory interpretation correspond to momentum and winding number, are connected with charges $q_1$ and $q_2$, corresponding to $U(1)$ gauge fields $F^{(1)}$ and $F^{(2)}$ (and defined by standard use of Sen's entropy function formalism), through the relations
\begin{equation} \label{chnorm}
q_1 = \frac{2n}{\sqrt{\alpha'}} \;\;, \qquad q_2 = \frac{2w}{\sqrt{\alpha'}} \;.
\end{equation}
Now, $v_2$ is the real positive root of an equation
\begin{equation} \label{v2eq}
\sum_{k=0}^{[D/2]-1} \frac{(D-2)!}{(D-2k-2)!\, k!} \left( \frac{\alpha'}{4\, v_2} \right)^{\!k} = 2
\end{equation}
which is a polynomial equation in $1/v_2$ of $([D/2]-1)$-th order. The equation can be analytically solved for $D\le9$. For $D=4$ and $D=5$ the solutions are given by
\begin{equation} \label{v2D45}
v_2 = \frac{\alpha'}{4}(D-2)(D-3)
\end{equation}
while for $D=6$ and $D=7$ they are given by
\begin{equation}
v_2 = \frac{\alpha'}{8}(D-2)(D-3) \left[ 1 + \sqrt{1 + \frac{2(D-4)(D-5)}{(D-2)(D-3)}} \right]
\end{equation}
We have not found analytic form of the solution for general $D$.

Sen's entropy function formalism allows one to find the black hole entropy as defined by Wald formula 
\cite{Iyer:1994ys}. The result, valid in all $D$, is \cite{Prester:2005qs}
\begin{equation} \label{SbhD}
S_{\mathrm{bh}} = 4\pi \sqrt{|nw|}
\end{equation}
which matches microscopic result (\ref{Smicro}) for BPS configurations. In fact, requirement of this matching fixes the coefficients $\lambda_k$ to the form (\ref{coup}) uniquely.\footnote{Note that the fact that action of the form (\ref{treeea}) can reproduce entropy formula (\ref{SbhD}) in all dimensions $D$ is nontrivial, because for every couple of dimensions $D$ one has only one coupling constant $\lambda_k$ to ``play with''.}

Lovelock-type theory (\ref{treeea})-(\ref{coup}) gives the same result for black hole entropy (\ref{SbhD}) for configurations with $nw<0$. This is not so in the heterotic string theory interpretation where 2-charge configurations with $nw<0$, which are not supersymmetric (non-BPS), have statistical entropy given by
\begin{equation} \label{SbhDII}
S_{\mathrm{micro}}^{(II)} = 2\sqrt{2}\pi \sqrt{|nw|} \;, \qquad |nw| \gg 1
\end{equation}
The same expression for the microscopic entropy one gets for corresponding 2-charge configurations with the elementary string in type-II superstring theories compactified on $S^1 \times T^{9-D}$, which are 1/4-BPS, regardless of the sign of $nw$. We note that the entropy 
(\ref{SbhDII}) can be reproduced by Lovelock-type action (\ref{treeea}) by taking the coefficients $\lambda_k$ to be
\begin{equation}\label{coupII}
\lambda_k^{(II)} = \frac{2\, \alpha'^{k-1}}{4^k k!} \,, 
\end{equation}
instead of (\ref{coup}). With this choice $AdS_2 \times S^{D-2}$ solutions can be obtained from those in (\ref{sollove}) through
\begin{equation} \label{solloveII}
v_{1,2}^{(II)} = \frac{v_{1,2}}{2} \;,\qquad e_{1,2}^{(II)} = \frac{e_{1,2}}{\sqrt{2}} \;,\qquad
 u_T^{(II)} = u_T \;,\qquad u_S^{(II)} = 2^{(D-3)/2} u_S \;.
\end{equation}
In the following we shall focus on solutions (\ref{sollove}). Using (\ref{solloveII}) one can easily extend all results to solutions of the theory with $\lambda_k$ given by (\ref{coupII}).\footnote{Nontrivial dependence of black hole entropy on sign of $nw$ can appear only if gravitational Chern-Simons terms are present in the action, and it is known that such terms appear in low-energy effective action of heterotic string theory but not of type-II theories (compactified on $S^1 \times T^{9-D}$). This statement can be made especially strong when near-horizon geometry is (locally) isometric to some geometry containing AdS$_3$ factor \cite{Kraus:2005vz,Kraus:2005zm}, which indeed typically happens in the heterotic large black hole solutions of low-energy effective actions. In view of this, let us emphasize that when we uplift our solutions (\ref{sollove}) and (\ref{solloveII}) in the standard way to $(D+1)$-dimensions, then the factor AdS$_2 \times S^1$ is \emph{not} locally isomorphic to AdS$_3$ (though this statement depends on the field-redefinition scheme used). This behavior was argued to happen in the type-II case in \cite{Dabholkar:2006tb}.}

\vspace{10pt}

\section{$D \to \infty$ limit}
\label{sec:Dtoinfty}

\bigskip

The simplicity of the near-horizon solution from the previous section makes it interesting toy model
for analyzing large $D$ limit. In particular, we see from (\ref{sollove}) that all dependence on $D$ is contained in parameter $v_2$, which is also the only parameter which is not trivial to find. On the other hand, in the $D \to \infty$ limit number of terms in the action becomes infinite (of the order $D/2$), which makes the limit a priori non-trivial (and also mimics behavior of effective string theory actions).

We take $D\to\infty$ while keeping other parameters ($G_D$, $\alpha'$) and charges ($n$ and $w$) fixed.In the leading order in large $D$ limit polynomial on the right-hand side of (\ref{v2eq}) can be summed
\begin{equation}
Q(v_2) \to \sum_{k=0}^\infty \frac{1}{m!} \left( \frac{D^2 \alpha'}{4\, v_2} \right)^{\!k}
 = \exp \left( \frac{D^2 \alpha'}{4\, v_2} \right)
\end{equation}
which means that equation (\ref{v2eq}) for $v_2$ takes the simple form in $D\to\infty$ limit
\begin{equation} \label{v2eqinf}
e^x = 2 \;, \qquad x \equiv \frac{D^2 \alpha'}{4\, v_2}
\end{equation}
That infinite series can be summed to give simple exponential function is a remarkable result, which may be connected to the conjecture from \cite{Prester:2005qs} that the gravitational part of the action (\ref{treeea})-(\ref{coup}) can be written in an exponential form (see Eq. (5.1) from 
\cite{Prester:2005qs}).

The solution of (\ref{v2eqinf}) is
\begin{equation} \label{v2inf}
v_2 \to \frac{D^2\, \alpha'}{4\, \ln2}
\end{equation}
Note that $v_2$ is the square of the proper radius of the $S^{D-2}$ sphere, so of the black hole horizon, by which we mean that a proper area of the horizon in the string frame metric is given by
\begin{equation} \label{areasfD}
A_{\mathrm{h}} = \Omega_{D-2}\, v_2^{(D-2)/2} \;\;, \qquad 
 \Omega_{D-2} = \frac{2 \pi^{(D-1)/2}}{\Gamma\left(\frac{D-1}{2}\right)}
\end{equation}
where $\Omega_{D-2}$ is the area of the standard $(D-2)$-sphere with unit radius. By putting 
$r_{\mathrm{h}}\equiv \sqrt{v_2}$ in (\ref{v2inf}) we obtain the relation (\ref{Dapexact}). Interestingly, the behavior $r_{\mathrm{h}} \sim D \sqrt{\alpha'}$ was previously obtained in \cite{Emparan:2013xia} by analyzing large $D$ limit of \emph{large} black holes in the ordinary general relativity, which is a rather different context in which parameter $\alpha'$ is not present but inferred through the comparison with string theory calculations.

Let us mention here that the property that $\alpha' v_1 \sim 1$ and $\alpha' v_2 \sim D^2$ is such that it guarantees that curvature scalars are generically finite and nonvanishing in $D \to \infty$ limit (i.e., they are $O(1)$ in $D$), which means that every Euler terms $E_k$, when evaluated on our solution, gives finite contribution in the Lagrangian density in $D \to \infty$ limit both in AdS$_2$ and $S^{D-2}$ block. This shows that in this theory there are some similarities between $1/D$ and $\alpha'$ expansions. We postpone more detailed discussion on this to the next section, where we shall also show that the things differ when one truncates higher-derivative part of the action to the fixed order.   

To complete large $D$ limit of the solution (\ref{sollove}) we have to calculate $u_S$. For this we need large $D$ behavior of $\Omega_{D-2}$ which is given by
\begin{equation} \label{Ominf}
\Omega_{D-2} \to \frac{D}{\pi \sqrt{2}} \left( \frac{2\pi e}{D} \right)^{D/2}
\end{equation}
Using (\ref{v2inf}) and (\ref{Ominf}) in expression for $u_S$ in (\ref{sollove}) we obtain
\begin{equation} \label{uSDinf}
u_S \,\to\, \frac{8\pi\, G_D}{e\, \sqrt{2}}\sqrt{|nw|} 
 \left( \frac{2 \ln2}{\pi e\, \alpha' D} \right)^{\frac{D}{2}-1}
\end{equation}
Together with (\ref{sollove}) and (\ref{v2inf}) this completes large $D$ limit of our solution.

In the fully quantized theory one expects that the field $\Phi$ is connected with the effective quantum coupling constant $g_{\mathrm{eff}}$ through a relation of the form
\begin{equation} \label{geffdef}
g_{\mathrm{eff}}^2 = g_0^2 \, e^\Phi = \frac{g_0^2}{u_S}
\end{equation}
From (\ref{uSDinf}) and (\ref{geffdef}) follows that if we want to have sensible theory in large $D$ limit, with meaningful perturbative expansion in which $g_{\mathrm{eff}}$ is finite and nonvanishing, we have to scale Newton constant $G_D$ and/or $\alpha'$ such that
\begin{equation} \label{GDg0alpha}
G_D \,\to\, g_0^2\, \frac{e\, \sqrt{2}}{8\pi\, b} \left( \zeta \alpha' D \right)^{\frac{D}{2}-1}
\end{equation}
where $\zeta$ and $b$ are $D$-independent dimensionless parameters.\footnote{In the string theory interpretation $\Phi$ is the dilaton field and the relation (\ref{geffdef}) can be written more explicitly for $D\le9$. However, it is not obvious what could define a valid extension to $D\to\infty$. For example, together with $D$ one can consider also letting number of compactified dimensions to infinity \cite{Canfora:2009dx} in some way. We note that $g_0$ is not constrained by our analysis, and in particular it may be $D$-dependent. These are interesting issues which however go beyond our 
tree-level analysis.} Then the effective coupling constant becomes
\begin{equation} \label{geffD}
g_{\mathrm{eff}}^2 \,\to\,
 \frac{b}{\sqrt{|nw|}} \left( \frac{\pi e}{\zeta\, 2 \ln2} \right)^{\frac{D}{2}-1}
\end{equation}
Now we see that there is a critical value of $\zeta$ defined by
\begin{equation} \label{zetac}
\zeta_c = \frac{\pi e}{2 \ln 2} = 6.16\ldots
\end{equation}
which separates the two phases in the $D\to\infty$ limit: $\zeta < \zeta_c$ for which $g_{\mathrm{eff}} \to 0$, and $\zeta > \zeta_c$ for which $g_{\mathrm{eff}} \to \infty$. In the critical point $\zeta = \zeta_c$ we obtain that $g_{\mathrm{eff}}$ is finite in $D\to\infty$ limit and
\begin{equation} \label{geffc}
g_{\mathrm{eff}}^2 \to \frac{b}{\sqrt{|nw|}}
\end{equation}
In the critical point the effective coupling constant is finite in $D \to \infty$ limit so there is a hope that in this case one can define sensible quantum theory. For us here it is important that from (\ref{geffc}) follows that $g_{\mathrm{eff}}$ can be made arbitrarily small by taking $|nw| \gg 1$ (which is the relevant regime in the string theory interpretation) which makes tree-level approximation credible. 

Let us say a few words more on the near-horizon geometry in $D\to\infty$ limit. Putting (\ref{Ominf}) and (\ref{v2inf}) in (\ref{areasfD}) one gets that in the ``string frame'' horizon area is
\begin{equation}
A_{\mathrm{h}} \,\to\, e\sqrt{2} \left( \frac{\pi e\, \alpha' D}{2 \ln2} \right)^{\frac{D}{2}-1}
 =\, e\sqrt{2}\, \left(\zeta_c\, \alpha' D \right)^{\frac{D}{2}-1}
\end{equation}
If we use the critical scaling (i.e., (\ref{GDg0alpha}) with $\zeta = \zeta_c$) we obtain
\begin{equation}
A_{\mathrm{h}} \,\to\, \frac{8\pi\, b}{g_0^2} G_D
\end{equation}
If $g_0$ is $D$-independent the dimensionless horizon area, when measured in the units of $G_D$, is finite and nonvanishing in $D\to\infty$ limit. 

For completeness, let us briefly analyze $D\to\infty$ limit of the geometry in the Einstein frame, in which metric is 
\begin{equation}
g_{\mu\nu}^{(\mathrm{E})} = u_S^{2/(D-2)} \, g_{\mu\nu}
\end{equation}
The horizon area in the Einstein frame is then
\begin{equation}
A_{\mathrm{h}}^{(\mathrm{E})} = u_S \, A_{\mathrm{h}} = 8\pi\,G_D \sqrt{|nw|}
\end{equation}
while for the square of the AdS$_2$ radius one gets
\begin{equation}
v_1^{(\mathrm{E})} = u_S^{\frac{2}{D-2}} \, v_1 \to \frac{\ln2}{\pi e D}\, G_D^{\frac{2}{D-2}}
\end{equation}
If we use again the critical scaling we obtain
\begin{equation}
v_1^{(\mathrm{E})} \to \frac{\alpha'}{2}\, g_0^{\frac{1}{D-2}}
\end{equation}
If $g_0$ is $D$-independent one obtains $v_1^{(\mathrm{E})} \to \alpha'/2 = v_1$. We see that in critical scaling, combined with an assumption that $g_0 \sim O(1)$ in $D$, leads to the same $O(1)$ behavior of geometries, both in string frame and Einstein frame, if one properly defines units of geometrical objects.

\vspace{10pt}

\section{Finite- vs. infinite- derivative action in $D \to \infty$ limit}
\label{sec:finite}

\bigskip

The natural question to ask is are the results presented in Sec. \ref{sec:Dtoinfty} special or general, i.e., does the choice of particular higher-derivative corrections, given by (\ref{treeea}) and (\ref{coup}) (or (\ref{coupII})), have some special consequences not shared by generic higher-derivative theories in $D\to\infty$. In particular, how important is the property that in $D\to\infty$ limit the Lagrangian of our Lovelock-type action has an infinite number of higher-derivative terms? We would ideally also  To throw some light one these issues, we take another simple example of a different kind - an action whose higher-derivative sector is terminated at 4-derivative (i.e., $R^2$) order and analyze its large $D$ limit. To achieve easy comparison with the results from the previous section, we take the new action to be
\begin{equation} \label{GBaction}
\mathcal{A} = \mathcal{A}_0 + 
 \frac{1}{16\pi G_D} \frac{\alpha'}{8} \int d^Dx \sqrt{-g} \,e^{-\Phi} \, E_2 \,,
\end{equation}
where $E_2$ is the second Euler density, known as the Gauss-Bonnet density, which by (\ref{lgbm}) is
\begin{equation} \label{GBterm}
E_2 = R_{\mu\nu\rho\sigma} R^{\mu\nu\rho\sigma} - 4\, R_{\mu\nu} R^{\mu\nu} + R^2
\end{equation}
and $\mathcal{A}_0$ is as in (\ref{eaDs}). This action is obtained from the action (\ref{treeea})-(\ref{coup}) by terminating the sum over higher-derivative terms in (\ref{treeea}) at the 4-derivative level (i.e., by keeping just the first element $k=2$ for all $D$). From now on we shall refer to this action as Gauss-Bonnet-type (GB-type) theory. 

Again, one can use Sen's entropy function formalism to find solutions of GB-type theory with AdS$_2 \times S^{D-2}$ geometry (\ref{nhsmall}) and the entropy of small extremal black holes with such near-horizon configurations. This calculation was already made in \cite{Sen:2005kj} and the results are:
\begin{eqnarray} \label{solGB}
&& v_1 = \frac{\alpha'}{2} \left[ 1 - \frac{(D-4)(D-5)}{(D-2)(D-3)} \right] \;\;, 
\qquad\qquad\qquad\;
   v_2 = \frac{\alpha'}{4} \left[(D-2)(D-3)-(D-4)(D-5)\right] \;,
\nonumber \\
&& u_S = \frac{16\pi G_D}{\Omega_{D-2}} \frac{v_1 \sqrt{|nw|}}{\alpha'\, v_2^{(D-2)/2}} 
 \left[ 1 - \frac{(D-4)(D-5)}{2(D-2)(D-3)} \right]^{-1/2} \;, 
\qquad\qquad
   u_T = \sqrt{\left| \frac{n}{w} \right|} \;,
\nonumber \\
&& e_1 = \frac{\sqrt{\alpha' |nw|}}{2n} \left[ 1 - \frac{(D-4)(D-5)}{2(D-2)(D-3)} \right]^{1/2} \;, \qquad
   e_2 = \frac{\sqrt{\alpha'|nw|}}{2w} \left[1-\frac{(D-4)(D-5)}{2(D-2)(D-3)}\right]^{1/2} \;.
\qquad\quad
\end{eqnarray}
where we again used the same (string theory motivated) normalization for electric charges (\ref{chnorm}). The black hole entropy is
\begin{equation} \label{entGB}
S_{\mathrm{bh}} = 4\pi \sqrt{|nw|} \left[1-\frac{(D-4)(D-5)}{2(D-2)(D-3)}\right]^{1/2}
\end{equation}
We see that for $D=4$ and $D=5$ (\ref{solGB})-(\ref{entGB}) is equal to (\ref{sollove}), 
(\ref{v2D45}) and (\ref{SbhD}), as it should. The simple 4-derivative Gauss-Bonnet term is also capable of regularizing small black holes in all $D$, though in $D>5$ dimensions it does not lead to black hole entropy which matches string theory statistical entropy (\ref{Smicro}). So, the GB-type theory, unlike Lovelock-type theory, does not have potential to offer effective description of 2-charge configurations in string theories when $D>5$.

We now perform the large $D$ limit. In the leading order the solution (\ref{solGB}) becomes
\begin{eqnarray} \label{solGBinf}
&& v_1 \to \frac{2 \alpha'}{D} \;\;, \qquad\quad v_2 \to \alpha' D \;\;, \qquad\qquad 
 u_S \to \frac{64\pi^2\, G_D\, \alpha' \sqrt{|nw|}}{D (2\pi e\, \alpha')^{D/2}} \;,
\nonumber \\
&& u_T = \sqrt{\left| \frac{n}{w} \right|} \;\;, \qquad
 e_1 \to \frac{\sqrt{\alpha' |nw|}}{2\sqrt{2}\,n} \;\;, \qquad
 e_2 \to \frac{\sqrt{\alpha'|nw|}}{2\sqrt{2}\,w}
\end{eqnarray}
while the black hole entropy is
\begin{equation}
S_{\mathrm{bh}} \to 2\sqrt{2}\pi \sqrt{|nw|}
\end{equation}
It is obvious that the $D$ dependence of the solution (\ref{solGBinf}) in GB-type theory is different from the one obtained in the Lovelock-type theory presented in Eqs. (\ref{sollove}), (\ref{v2inf}) and 
(\ref{uSDinf}). In particular, instead of the relation (\ref{Dapexact}) one now obtains
\begin{equation} \label{DapGB}
\sqrt{\alpha'} = \frac{r_{\mathrm{h}}}{\sqrt{D}} \;,
\end{equation}
where again $r_{\mathrm{h}} = \sqrt{v_2}$ is the proper radius of the horizon. The relation (\ref{DapGB}) has different scaling in $D$ from (\ref{Daprime}). So, our first conclusion is that the scaling (\ref{Daprime}) is not a generic result (in the setup we consider), and that the scaling depends crucially on the properties of the higher-derivative sector.

Let us now analyze the geometry in GB-type theory more closely. In the string frame the area of $(D-2)$-sphere (horizon area) is now
\begin{equation} \label{geomGB}
A_{\mathrm{h}} \to \sqrt{2} e\, (2\pi e\, \alpha')^{D/2-1}
\end{equation}
In the Einstein frame we obtain that the squared radius of AdS$_2$ factor and proper area of $(D-2)$-sphere  are
\begin{equation} \label{geomEGB}
v_1^{(E)} \to \frac{G_D^{2/(D-2)}}{\pi e\, D} \;\;, \qquad
 A_{\mathrm{h}}^{(\mathrm{E})} \to 32\pi\, G_D \sqrt{2|nw|} \frac{1}{D}
\end{equation}
Again, one obtains completely different $D$-dependence compared to those in Lovelock-type theory. In particular, we note that factors of the type $D^{D/2}$ are completely absent in (\ref{solGBinf})-(\ref{geomEGB}), if they are not introduced through parameters ($G_D$ and $\alpha'$) or charges ($n$ and $w$).

From the expression for $u_S$ in (\ref{solGBinf}) we obtain that the analogue of the critical scaling condition (\ref{GDg0alpha}) and (\ref{zetac}) is now
\begin{equation}
G_D \,\to\, g_0^2\, \frac{b\, e}{32\pi} D (2\pi e\, \alpha')^{\frac{D}{2}-1} \;\;,
\end{equation}
In the critical scaling the horizon area, both in string frame and Einstein frame, are finite in $D\to\infty$ limit when measured in units of $(2\pi e\,\alpha')$.

A mayor difference between solutions in Lovelock-type and GB-type actions is in the $D\to\infty$ limit of curvature scalars. To show this, first not that all irreducible curvature scalars (those which are not products of two scalars) for AdS$_2 \times S^{D-2}$ geometry are sum of AdS$_2$ contribution (denoted with subscript $A$) and $S^{D-2}$ contribution (denoted with the subscript $S$), e.g.,
\begin{equation}
R = R_A + R_S \;\;, \qquad (R_{\mu\nu})^2 = (R_{\mu\nu})_A^2 + (R_{\mu\nu})_S^2 \;\;, \quad 
\mathrm{etc.}
\end{equation}
In the Lovelock-type theory (\ref{treeea}) generic curvature scalars are $O(1)$ in $D$. More precisely, all AdS$_2$ scalars are $D$-independent (because $v_1 = \alpha'/2$), while of $S^{D-2}$ irreducible scalars only 
$R_S \sim O(1)$ while all others are subleading in $1/D$, e.g.,
\begin{equation}
\alpha'^2 (R_{\mu\nu})_S^2 \to \frac{1}{D} \;\;, \qquad  
\alpha'^2 (R_{\mu\nu\rho\sigma})_S^2 \to \frac{2}{D^2} \;\;.
\end{equation}
As a consequence every Euler term in the Lovelock-type action produces a finite contribution in 
$D \to \infty$ limit. We can be even more precise. Already for finite $D$ we know that AdS$_2$ contribution of curvature terms to equations of motion is solely through $R_A$ (and linearly). When combined with previously said, we obtain that in the leading order of large $D$ limit contributions from gravity sector to AdS$_2 \times S^{D-2}$ solutions are just factors of Ricci scalars $R_A$ and $R_S$, which are linear in $R_A$. 

In the GB-type theory (\ref{GBaction}) one obtains in the large $D$ limit the following expressions
\begin{eqnarray}
&& \alpha' R_A \to - D \;\;,\qquad\, \alpha'^2 (R_{\mu\nu})_A^2 \to \frac{D^2}{2} \;\;, \qquad\,
 \alpha'^2 (R_{\mu\nu\rho\sigma})_A^2 \to D^2 \;\;,
\nonumber \\
&& \alpha' R_S \to D \;\;,\qquad\quad \alpha'^2 (R_{\mu\nu})_S^2 \to D \;\;, \qquad\quad
 \alpha'^2 (R_{\mu\nu\rho\sigma})_S^2 \to 2 \;\;.
\end{eqnarray}
We see that here the leading order in $D$ comes \emph{solely} from the 4-derivative terms (which are highest-derivative terms) and that curvature scalars are of the order $D^2$, which is completely different behavior then in the Lovelock-type theory.\footnote{Note that in both theories the $S$-block contributes only through Ricci scalar factors $R_S$ in the leading order for large $D$.}

From the analysis of this particular GB-type theory some generic conclusions can be drawn. One is that infinite number of terms (of the order $D$) with unlimited number of derivatives (of the order $D/2$) present in the Lovelock-type action is responsible for markedly different large $D$ behavior of solutions, compared with the theories which have fixed order in derivatives. However, our analysis does not reveal are there some properties of $D\to\infty$ limit which are special for our Lovelock-type action, when applied to AdS$_2 \times ^{D-2}$ configurations. We plan to investigate this in the future.

\vspace{20pt}

\noindent
{\bf \large Acknowledgements}

\bigskip

\noindent
We thank Zdravko Lenac for stimulating discussions. The research was supported by the Croatian Ministry of Science, Education and Sport under the contract no.~119-0982930-1016.

\vspace{30pt}


\end{document}